\documentclass[11pt,a4paper]{article}
\usepackage{amssymb}
\usepackage{amsmath}
\usepackage{amsthm}
\usepackage{parskip}
\usepackage[margin=1in]{geometry}

\title{Improved Approximation of Infinite Thermostat by Finite Reservoir Using the 3D Kac Model}
\author{Federico Bonetto, Anthony Popa, Matthew Powell, Peter Chen, Steven Tung\\
\\
Georgia Institute of Technology, School of Mathematics,\\
Atlanta, Georgia, 30332, USA}
\date{\today}

\begin{document}

\maketitle

\begin{abstract}
In this paper, we study a system of $M$ particles interacting with a reservoir of $N$ particles, where $N >> M$, and compare this setup to one where the $M$-particle system interacts with a thermostat of infinite particles. Our goal is to prove a suitable upper bound, uniform in time, on the distance between the states of these two setups, given an initial Maxwellian state for both the reservoir and thermostat. Previous work has analyzed this problem using the one-dimensional Kac Model of gas collisions and an $L^2$ norm to define distance; the result was a bound which scaled with $M/\sqrt{N}$. In this paper, we use the $L^2$ norm and the three-dimensional generalization of the Kac Model to prove a bound whose long-term behavior scales with $M/N$.
\end{abstract}

\section{Introduction}

In 1956, Marc Kac introduced a simplified, one-dimensional model of gas collisions known as the Kac Model, in which collisions have a fixed likelihood of occurring at any given instant, and all collisions which preserve energy are equally valid. The original purpose of the model was to prove that Boltzmann's chaotic assumption propagates in time. \cite{kac} 

In recent years, however, the Kac Model has been used for another purpose; studying approach to equilibrium for thermodynamic systems. Due to its simplicity, it allows us to explicitly perform computations which would otherwise be intractable, while still having enough richness as a model to capture the essential behavior of systems. 

In 2017, Bonetto et. al. used the Kac Model to compare two setups: an $M$-particle system interacting with an infinite thermostat of particles, and an $M$-particle system interacting with a finite $N$-particle reservoir, where $N >> M$. Using a suitable $L^2$ norm, they proved an upper bound on the distance between the states of these two setups that scales with $M/\sqrt{N}$. \cite{bonetto} In this paper, using the same norm, we prove a uniform bound which includes a $M/\sqrt{N}$ factor for small time but scales with $M/N$ as time approaches infinity.

This paper also makes use of a 3D generalization of the Kac Model, which is similar to the original 1D version except we only allow collisions which preserve both energy and momentum, as opposed to the 1D model where momentum is in general not preserved.

\section{Definitions and Initial Derivations}

The velocity vector of the $M$-particle system is defined as $\mathbf{v} = (\vec{v}_1, \dots, \vec{v}_M)$, with $\vec{v}_i = (v_{i,1}, v_{i,2}, v_{i,3})$. The state of the system is given by the probability density function $f(\mathbf{v})$.

If two particles $i$ and $j$ collide with resulting system velocity $\mathbf{v}$, the pre-collision system velocity is given by:
\[
\mathbf{v}_{ij}^*(\vec{\Omega}) = (\vec{v}_1, \dots, \vec{v}_i^*, \dots, \vec{v}_j^*, \dots, \vec{v}_M),
\]

\[
\vec{v}_i^* = \vec{v}_i - \big[ (\vec{v}_i - \vec{v}_j) \cdot \vec{\Omega} \big] \vec{\Omega},
\quad \vec{v}_j^* = \vec{v}_j - \big[ (\vec{v}_j - \vec{v}_i) \cdot \vec{\Omega} \big] \vec{\Omega}
\]
where $\vec{\Omega}$ is a unit vector representing the relative orientation of particles $i$ and $j$.

According to the 3D Kac Model, the infinitesimal generator for the evolution of $f$ is given by:
\[
\mathcal{L}_S[f] 
= \frac{\lambda_S}{M-1} \sum_{1 \le i < j \le M} \big( R_{ij}^S[f] - f \big),
\]

\[
R_{ij}^S[f](\mathbf{v}) 
= \int_{\mathbb{S}^2} f\big(\mathbf{v}_{ij}^*(\vec{\Omega})\big) \, d\sigma(\vec{\Omega})
\]
Here, $\lambda_S dt$ is the probability for a specific particle to collide with any of the other particles in a time interval $dt$. $R_{ij}^S$ is an average of $f$ over all possible pre-collision velocities; $d\sigma(\vec{\Omega})$ is the normalized Haar measure on $\mathbb{S}^2$.

We now introduce an interaction between the $M$-particle system and a thermostat, an infinite sea of particles whose velocities have a Gaussian probability distribution. This generator is given by:
\[
\mathcal{L}_B[f] = \mu \sum_{i=1}^M \big( B_i[f] - f \big),
\]
\[
B_i[f](\mathbf{v}) 
= \int_{\mathbb{R}^3} \int_{\mathbb{S}^{2}} 
e^{-\pi |\vec{x}^*(\vec{v}_i, \vec{\Omega})|^2 } 
\, f\big(\mathbf{v}_{i}^*(\vec{x}, \vec{\Omega}))\big) 
\, d\sigma(\vec{\Omega}) \, d\vec{x}
\]
$B_i$ is the gain from a collision between system particle $i$ and a thermostat particle. It averages $f$ over all possible pre-collision velocities of the system particle, weighted by the probability of the thermostat particle having the corresponding pre-collision velocity. The velocity vectors are dealt with similarly to the case of internal system collisions:
\[
\mathbf{v}_{i}^*(\vec{x}, \vec{\Omega}) = (\vec{v}_1, \dots, \vec{v}_i^*, \dots,\vec{v}_M),
\]

\[
\vec{v}_i^* = \vec{v}_i - \big[ (\vec{v}_i - \vec{x}) \cdot \vec{\Omega} \big] \vec{\Omega},
\quad
\vec{x}^*(\vec{v}_i, \vec{\Omega}) = \vec{x} - \big[ (\vec{x} - \vec{v}_i) \cdot \vec{\Omega} \big] \vec{\Omega}
\]

Analyzing the evolution of the system over time, we are interested in how closely the effect of the thermostat matches the effect of a finite reservoir of $N$ particles, where $N >> M$. From now on, we call the combination of system and thermostat the T-system, and the combination of system and reservoir the R-system.

Because the thermostat is infinite, the $M$-particle system will not change the thermostat's state. But the reservoir's state will change when interacting with the system. We define the reservoir's $N$-particle velocity vector as:
\[
\mathbf{w} = (\vec{w}_1, \dots, \vec{w}_N), 
\quad \vec{w}_i = (w_{i,1}, w_{i,2}, w_{i,3})
\]
When discussing the R-system, we now use a probability density function for both the system and reservoir:
\[
f(\mathbf{v}, \mathbf{w})
\]

Evolution of the system under internal collisions is now given by:
\[
R_{ij}^S[f](\mathbf{v}, \mathbf{w}) 
= \int_{\mathbb{S}^2} f\big(\mathbf{v}_{ij}^*(\vec{\Omega}), \mathbf{w}\big) \, d\sigma(\vec{\Omega})
\]

For internal collisions in the reservoir, we have:
\[
\mathbf{w}_{ij}^*(\vec{\Omega}) = (\vec{w}_1, \dots, \vec{w}_i^*, \dots, \vec{w}_j^*, \dots, \vec{w}_N),
\]

\[
\vec{w}_i^* = \vec{w}_i - \big[ (\vec{w}_i - \vec{w}_j) \cdot \vec{\Omega} \big] \vec{\Omega},
\quad
\vec{w}_j^* = \vec{w}_j - \big[ (\vec{w}_j - \vec{w}_i) \cdot \vec{\Omega} \big] \vec{\Omega}
\]

The corresponding generator is:
\[
\mathcal{L}_R[f] 
= \frac{\lambda_R}{N-1} \sum_{1 \le i < j \le N} \big( R_{ij}^R[f] - f \big),
\]
\[
R_{ij}^R[f](\mathbf{v}, \mathbf{w}) 
= \int_{\mathbb{S}^2} f\big(\mathbf{v}, \mathbf{w}_{ij}^*(\vec{\Omega})\big) \, d\sigma(\vec{\Omega})
\]

Finally to complete the R-system evolution, we need to consider collisions of system particles with reservoir particles. The pre-collision velocity vectors are:
\[
\mathbf{v}_{i}^*(\vec{\Omega}) = (\vec{v}_1, \dots, \vec{v}_i^*, \dots,\vec{v}_M),
\quad
\mathbf{w}_{j}^*(\vec{\Omega}) = (\vec{w}_1, \dots, \vec{w}_j^*, \dots, \vec{w}_N),
\]

\[
\vec{v}_i^* = \vec{v}_i - \big[ (\vec{v}_i - \vec{w}_j) \cdot \vec{\Omega} \big] \vec{\Omega},
\quad
\vec{w}_j^* = \vec{w}_j - \big[ (\vec{w}_j - \vec{v}_i) \cdot \vec{\Omega} \big] \vec{\Omega}
\]

The interaction generator is defined as:
\[
\mathcal{L}_I[f] 
= \frac{\mu}{N} \sum_{i=1}^M \sum_{j=1}^N \big( R_{ij}^I[f] - f \big)
\]
\[
R_{ij}^I[f](\mathbf{v}, \mathbf{w}) 
= \int_{\mathbb{S}^2} f\big(\mathbf{v}_{i}^*(\vec{\Omega}), \mathbf{w}_{j}^*(\vec{\Omega})\big) \, d\sigma(\vec{\Omega})
\]
Here, $\mu dt$ is the probability that a specific particle in the system collides with any particle in the reservoir.

Putting it all together, the infinitesimal generator for the evolution of the R-system is:
\[
\mathcal{L} = \mathcal{L}_S + \mathcal{L}_R + \mathcal{L}_I
\]

In the T-system, there is no reservoir involved. However, for ease of computation, we will use the joint probability density function from the R-system to describe the state of the T-system as well. This means the reservoir will be treated as having only internal collisions and will not affect the T-system interaction. In this style, the thermostat interaction generator $\mathcal{L}_B$ is the same, except now we have:

\[
B_i[f](\mathbf{v}, \mathbf{w}) 
= \int_{\mathbb{R}^3} \int_{\mathbb{S}^{2}} 
e^{-\pi |\vec{x}^*(\vec{v}_i, \vec{\Omega})|^2 } 
\, f\big(\mathbf{v}_{i}^*(\vec{x}, \vec{\Omega}), \mathbf{w})\big) 
\, d\sigma(\vec{\Omega}) \, d\vec{x}
\]

The generator for the T-system is thus:
\[
\tilde{\mathcal{L}} = \mathcal{L}_S + \mathcal{L}_R + \mathcal{L}_B
\]

We define:
\[
\Gamma(\vec{x}) = e^{-\pi|\vec{x}|^2}, 
\quad 
\Gamma(\mathbf{x^1},...,\mathbf{x^n}) = \prod_{i=1}^{K_1} \Gamma(\vec{x}^1_i)\cdots \prod_{i=1}^{K_n} \Gamma(\vec{x}^n_i).
\]
for a set of $n$ vectors $(\mathbf{x^1},...,\mathbf{x^n})$ in $(\mathbb{R}^{3K_1},...,\mathbb{R}^{3K_n})$.

The initial system-reservoir state (for both evolutions) is given by $f_0(\mathbf{v},\mathbf{w})$, where$\int_{\mathbb{R}^{3(M+N)}} f_0(\mathbf{v},\mathbf{w}) d\mathbf{v} d\mathbf{w} = 1$. The states of the R-system and T-system over time are given by:
\[
f_t = e^{\mathcal{L}t} f_0,
\quad
\tilde{f_t} = e^{\tilde{\mathcal{L}}t} f_0
\]

We set the reservoir to begin in a Maxwellian state like the thermostat, and write:
\[
f_0(\mathbf{v},\mathbf{w}) = h_0(\mathbf{v})\, \Gamma(\mathbf{v},\mathbf{w})
\]
We want to represent the evolution in terms of $h_0$ rather than $f_0$; this will enable us to define a convenient $L^2$ space to work with. Plugging $f_0$ into our rotational averaging operators, we see
\[
R_{ij}^S[f_0](\mathbf{v},\mathbf{w}) 
= \int_{\mathbb{S}^{2}} h_0\big(\mathbf{v}_{ij}^*(\vec{\Omega})\big) \Gamma(\mathbf{v}_{ij}^*(\vec{\Omega}),\mathbf{w}) \, d\sigma(\vec{\Omega})
\]

\[
= \int_{\mathbb{S}^{2}} h_0\big(\mathbf{v}_{ij}^*(\vec{\Omega})\big) \Gamma(\mathbf{v},\mathbf{w}) \, d\sigma(\vec{\Omega})
\]

\[
= \Gamma(\mathbf{v},\mathbf{w})\, R_{ij}^S[h_0]
\]
and likewise for $R_{ij}^R$ and $R_{ij}^I$. Therefore, we can write:
\[
f_t(\mathbf{v},\mathbf{w}) = h_t(\mathbf{v}, \mathbf{w})\, \Gamma(\mathbf{v},\mathbf{w}),
\]

\[
h_t = e^{\mathcal{L}t} h_0
\]

For the T-system, it is a little more complex. When we plug $f_0$ into $B_i$, we get
\[
B_i[f_0](\mathbf{v}, \mathbf{w}) 
= \int_{\mathbb{R}^3} \int_{\mathbb{S}^{2}} 
e^{-\pi |\vec{x}^*(\vec{v}_i, \vec{\Omega})|^2 } 
\, h_0\big(\mathbf{v}_{i}^*(\vec{x}, \vec{\Omega})\big)
\, \Gamma\big(\mathbf{v}_{i}^*(\vec{x}, \vec{\Omega}), \mathbf{w}\big) 
\, d\sigma(\vec{\Omega}) \, d\vec{x}
\]

\[
= \int_{\mathbb{R}^3} \int_{\mathbb{S}^{2}} 
\Gamma({\vec{x}})
\, h_0\big(\mathbf{v}_{i}^*(\vec{x}, \vec{\Omega})\big)
\, \Gamma\big(\mathbf{v}, \mathbf{w}) 
\, d\sigma(\vec{\Omega}) \, d\vec{x}
\]

\[
= \Gamma(\mathbf{v},\mathbf{w})\, T_i[h_0](\mathbf{v}),
\]
where we define:
\[
T_i[h](\mathbf{v})
= \int_{\mathbb{R}^3} \int_{\mathbb{S}^{2}} 
\Gamma({\vec{x}})
\, h\big(\mathbf{v}_{i}^*(\vec{x}, \vec{\Omega})\big)
\, d\sigma(\vec{\Omega}) \, d\vec{x}
\]

Thus:
\[
\tilde{f_t}(\mathbf{v},\mathbf{w}) = \tilde{h_t}(\mathbf{v})\, \Gamma(\mathbf{v},\mathbf{w}),
\]

\[
\tilde{h_t} = e^{\bar{\mathcal{L}}t} h_0,
\]

\[
\bar{\mathcal{L}} = \mathcal{L}_S + \mathcal{L}_R + \mathcal{L}_T,
\]

\[
\mathcal{L}_T = \mu \sum_{i=1}^M \big( T_i[f] - f \big)
\]

We define the $L^2(\mathbb{R}^{3(M+N)},\Gamma)$ space with the following inner product and norm:
\[
\langle f, g \rangle 
= \int_{\mathbb{R}^{3(M+N)}} f(\mathbf{v},\mathbf{w}) g(\mathbf{v},\mathbf{w}) \, \Gamma(\mathbf{v},\mathbf{w}) \, d\mathbf{v} \, d\mathbf{w}
\]

\[
\| f \| = \sqrt{ \langle f, f \rangle }
\]

$h_0$ is a function in this space, with
\[
\langle h_0, 1 \rangle = \int_{\mathbb{R}^{3(M+N)}} f_0(\mathbf{v},\mathbf{w}) d\mathbf{v} d\mathbf{w} = 1
\]

It will be convenient to associate functions which depend only on $\mathbf{v}$ with the subspace $L^2(\mathbb{R}^{3M},\Gamma)$, defined with inner product:
\[
\langle f, g \rangle 
= \int_{\mathbb{R}^{3M}} f(\mathbf{v}) g(\mathbf{v}) \, \Gamma(\mathbf{v}) \, d\mathbf{v}
\]

$\mathcal{L}$ and $\bar{\mathcal{L}}$ are bounded, self-adjoint operators on the space, and our goal is to find a bound for
\[
\|h_t - \tilde{h}_t\| = \|\big( e^{\mathcal{L}t} - e^{\bar{\mathcal{L}}t} \big) h_0\|
\]

One more ingredient which will be essential in our proof is the operator $\mathbf{R}$, which we define below as:
\[
\mathbf{R}[h](\mathbf{v},\mathbf{w}) = \int_{\mathcal{O}} h(O[\mathbf{v},\mathbf{w}]) \, d\sigma(O),
\]
where $\mathcal{O}$ is the set of all transformations preserving the total energy and momentum of the system and reservoir at velocities $\mathbf{v}$, $\mathbf{w}$. More precisely, we define three orthonormal vectors $\mathbf{g_1}$, $\mathbf{g_2}$, and $\mathbf{g_3}$ in $\mathbb{R}^{3(M+N)}$:
{\scriptsize
\[
\mathbf{g_1} = \frac{1}{\sqrt{M+N}}((1,0,0), ...,(1,0,0))\quad
\mathbf{g_2} = \frac{1}{\sqrt{M+N}}((0,1,0), ...,(0,1,0))\quad
\mathbf{g_3} = \frac{1}{\sqrt{M}+N}((0,0,1), ...,(0,0,1))
\]
}
$\mathcal{O}$ is defined as the subset of $SO(3(M+N))$ which preserves $\mathbf{g_i}$, $i=1,2,3$. $d\sigma(O)$ is the normalized Haar measure on $\mathcal{O}$. 

Put in words, $\mathbf{R}[h](\mathbf{v},\mathbf{w})$ is the average of $h$ over all valid system and reservoir velocity vectors with the same total energy and momentum as $(\mathbf{v},\mathbf{w})$. Note that $\mathbf{R}$ is an orthogonal projector onto the space of functions which are invariant under momentum-preserving rotations.

\section{Auxiliary Results}

\textbf{Lemma 1.}
For any function $h(\mathbf{v})$ in $L^2(\mathbb{R}^{3M},\Gamma)$ satisfying $\langle h, 1 \rangle = 1$,
\[
\|\mathbf{R}[h] - 1\| \le  C\|h - 1\|, \quad \text{where } C = \sqrt{\frac{3M}{3N-5}} + \sqrt{\left(\frac{M+N}{N}\right)^3 - 1}
\]
Note that expanding the expression inside the second square root gives $(M/N)^3 + 3(M/N)^2 + 3(M/N)$, so $C$ scales with $M/N$.

\begin{proof}

We begin the proof by defining orthonormal vectors to form a basis for $\mathbb{R}^{3(M+N)}$. Firstly, in keeping with the definition of $\mathbf{g_1}$, $\mathbf{g_2}$, and $\mathbf{g_3}$ in the previous section, for $i=1,2,3$:
\[
\mathbf{g_i} = \frac{1}{\sqrt{M+N}}(\sqrt{M}\mathbf{e_i}, \sqrt{N}\mathbf{f_i}),
\]
with $\mathbb{R}^{3M}$ vectors $\mathbf{e_i}$ defined as:

{\footnotesize
\[
\mathbf{e_1} = \frac{1}{\sqrt{M}}((1,0,0), ...,(1,0,0))\quad
\mathbf{e_2} = \frac{1}{\sqrt{M}}((0,1,0), ...,(0,1,0))\quad
\mathbf{e_3} = \frac{1}{\sqrt{M}}((0,0,1), ...,(0,0,1))
\]
}
and $\mathbb{R}^{3N}$ vectors $\mathbf{f_i}$:
{\footnotesize
\[
\mathbf{f_1} = \frac{1}{\sqrt{N}}((1,0,0), ...,(1,0,0))\quad
\mathbf{f_2} = \frac{1}{\sqrt{N}}((0,1,0), ...,(0,1,0))\quad
\mathbf{f_3} = \frac{1}{\sqrt{N}}((0,0,1), ...,(0,0,1))
\]
}

We define three more vectors $\mathbf{l_i}$ orthonormal to $\mathbf{g_i}$ as:
\[
\mathbf{l_i} = \frac{1}{\sqrt{M+N}}(\sqrt{N}\mathbf{e_i}, -\sqrt{M}\mathbf{f_i})
\]

Taking $\mathbf{e_i}$, we use the Gram-Schmidt process to find $3M-3$ vectors $\mathbf{a_1}$,...,$\mathbf{a_{3M-3}}$ to complete a basis for $\mathbb{R}^{3M}$. We do the same thing for $\mathbf{f_i}$, completing a basis for $\mathbb{R}^{3N}$ with $3N-3$ vectors $\mathbf{b_1}$,...,$\mathbf{b_{3N-3}}$. 

Thus, we have an orthonormal basis for $\mathbb{R}^{3(M+N)}$, and we define the orthogonal matrix $P$ as:
\[
P = [\mathbf{a_1},...,\mathbf{a_{3M-3}},\mathbf{g_1}, \mathbf{g_2}, \mathbf{g_3}, \mathbf{l_1}, \mathbf{l_2}, \mathbf{l_3},\mathbf{b_1},...,\mathbf{b_{3M-3}}],
\]
where the $\mathbf{a}$ vectors are padded with zeroes for entries $3M+1$ to $3(M+N)$, and the $\mathbf{b}$'s are padded with zeroes for entries 1 to $3M$.

$P^T$ maps the vector $(\mathbf{v}, \mathbf{w})$ to a vector we will denote as $(\mathbf{u}, \vec{V}, \vec{Y}, \mathbf{x})$, where $\mathbf{u}$ is in $\mathbb{R}^{3M-3}$, $\mathbf{x}$ is in $\mathbb{R}^{3N-3}$, and $\vec{V}$ and $\vec{Y}$ are in $\mathbb{R}^{3}$. Importantly, $\vec{V} = \frac{1}{M+N}\vec{p}$, where $\vec{p}$ is the R-system momentum.

We use $P$ to rewrite $\mathbf{R}$:
\[
\mathbf{R}[h](\mathbf{v},\mathbf{w}) = \int_{\mathcal{O}} h(O[\mathbf{v},\mathbf{w}]) \, d\sigma(O)
\]
\[
= \int_{\mathcal{O'}} h(PO'P^T[\mathbf{v},\mathbf{w}]) \, d\sigma(O') = \int_{\mathcal{O'}} h(PO'[\mathbf{u},\vec{V}, \vec{Y},\mathbf{x}]) \, d\sigma(O')
\]

$\mathcal{O'}$ is the subset of all rotations of $(\mathbf{u},\vec{V}, \vec{Y},\mathbf{x})$ which preserve the value of $\vec{V}$. Because length is preserved, we define $r^2 = |\mathbf{v}|^2 + |\mathbf{w}|^2 - |\vec{V}|^2 = |\mathbf{u}|^2 + |\vec{Y}|^2+ |\mathbf{x}|^2$, and write:
\[
\mathbf{R}[h](\mathbf{v},\mathbf{w}) = \int_{\mathbb{S}^{3(M+N)-4}(r)} h(P[\mathbf{u}',\vec{V}, \vec{Y}',\mathbf{x}']) \, d\sigma_r(\mathbf{u}', \vec{Y}',\mathbf{x}')
\]

Remember that $h$ depends only on $\mathbf{v}$, so it is a function of only the first $3M$ variables in its argument. We also know that, due to the structure of $P$, $\mathbf{x}'$ can have no affect on the first $3M$ variables, so we can write $h(P[\mathbf{u}',\vec{V}, \vec{Y}',\mathbf{x}']) = h(\tilde{P}[\mathbf{u}',\vec{V}, \vec{Y}'])$, where
\[
\tilde{P} = [\mathbf{a_1},...,\mathbf{a_{3M-3}},\mathbf{g_1}, \mathbf{g_2}, \mathbf{g_3}, \mathbf{l_1}, \mathbf{l_2}, \mathbf{l_3}]
\]

Therefore,
\[
\mathbf{R}[h](\mathbf{v},\mathbf{w}) = \int_{\mathbb{S}^{3(M+N)-4}(r)} h(\tilde{P}[\mathbf{u}',\vec{V}, \vec{Y}']) \, d\sigma_r(\mathbf{u}', \vec{Y}',\mathbf{x}')
\]
\[
= \frac{|\mathbb{S}^{3N-4}|}{|\mathbb{S}^{3(M+N)-4}|r^{3M}} \int_{\mathbb{R}^{3M}} \bar{h}(\vec{V}
,\mathbf{t}) \left(1-\frac{\mathbf{|t|}^2}{r^2}\right)_{+}^{\frac{3N-5}{2}} d\mathbf{t},
\]
where for simplicity we define $\mathbf{t} = [\mathbf{u}',\vec{Y}']$ and $\bar{h}(\vec{V},\mathbf{t}) = h(\tilde{P}[\mathbf{u}',\vec{V}, \vec{Y}'])$. The subscript $+$ is defined such that $(x)_{+} = x$ if $x \ge 0$ and $(x)_{+} = 0$ otherwise.

Our goal is to compute $\|\mathbf{R}[h] - 1\|$. In order to do this, we introduce the function $H$:
\[
H(\vec{V}) = \int_{\mathbb{R}^{3M}} \bar{h}(\vec{V}, \mathbf{t}) \, \Gamma({\mathbf{t}}) \, d\mathbf{t}
\]

Invoking the triangle inequality, we write:
\[
\|\mathbf{R}[h] - 1\| \le \|\mathbf{R}[h] - H\| + \|H - 1\|
\]
Our task is now to find bounds for both terms.

Starting with the first term, we write:
\[
\|\mathbf{R}[h] - H\|^2 = \int_{\mathbb{R}^{3(M+N)}} (\mathbf{R}[h] - H)^2 \, \Gamma(\mathbf{v}, \mathbf{w}) \, d\mathbf{v} \, d\mathbf{w}
\]
\[
= \int_{0}^{\infty} \int_{\mathbb{R}^{3}} |\mathbb{S}^{3(M+N)-4}|\, r^{3(M+N)-4} \, (\mathbf{R}[h] - H)^2 \, \Gamma(r, \vec{V}) \, d\vec{V} \, dr
\]

Since $\int_{\mathbb{R}^{3M}}  \, \Gamma({\mathbf{t}}) \, d\mathbf{t}$ = 1 and 
\[
\frac{|\mathbb{S}^{3N-4}|}{|\mathbb{S}^{3(M+N)-4}|r^{3M}} \int_{\mathbb{R}^{3M}}  \left(1-\frac{\mathbf{|t|}^2}{r^2}\right)_{+}^{\frac{3N-5}{2}} d\mathbf{t} = 1,
\]
we write:
{\footnotesize
\[
\mathbf{R}[h](\mathbf{v},\mathbf{w}) - H(\vec{V}) = \int_{\mathbb{R}^{3M}} \left[\frac{|\mathbb{S}^{3N-4}|}{|\mathbb{S}^{3(M+N)-4}|r^{3M}} \left(1-\frac{\mathbf{|t|}^2}{r^2}\right)_{+}^{\frac{3N-5}{2}} - \Gamma(\mathbf{t})\right] (\bar{h}(\vec{V}
,\mathbf{t}) - 1) \, d\mathbf{t}
\]
}
{\footnotesize
\[
= \int_{\mathbb{R}^{3M}} \left[\frac{|\mathbb{S}^{3N-4}|}{|\mathbb{S}^{3(M+N)-4}|r^{3M}} \left(1-\frac{\mathbf{|t|}^2}{r^2}\right)_{+}^{\frac{3N-5}{2}}\Gamma^{-1/2}(\mathbf{t}) - \Gamma^{1/2}(\mathbf{t})\right] \Gamma^{1/2}(\mathbf{t})(\bar{h}(\vec{V}
,\mathbf{t}) - 1) \, d\mathbf{t}
\]
}

Invoking the Cauchy-Schwarz inequality:
\[
(\mathbf{R}[h] - H)^2 \le \int_{\mathbb{R}^{3M}} \Gamma(\mathbf{t})(\bar{h}(\vec{V}
,\mathbf{t}) - 1)^2 d\mathbf{t}
\]
\[
\times \int_{\mathbb{R}^{3M}} \left[\frac{|\mathbb{S}^{3N-4}|}{|\mathbb{S}^{3(M+N)-4}|r^{3M}} \left(1-\frac{\mathbf{|t|}^2}{r^2}\right)_{+}^{\frac{3N-5}{2}}\Gamma^{-1/2}(\mathbf{t}) - \Gamma^{1/2}(\mathbf{t})\right]^2 d\mathbf{t}
\]

Plugging this inequality into our expression for $\|\mathbf{R}[h] - H\|^2$, we get:
\[
\|\mathbf{R}[h] - H\|^2 \le \int_{\mathbb{R}^3}\int_{\mathbb{R}^{3M}} \Gamma(\mathbf{t})(\bar{h}(\vec{V}
,\mathbf{t}) - 1)^2 d\mathbf{t} d\vec{V}
\]
{\footnotesize
\[
\times \int_{0}^{\infty} |\mathbb{S}^{3(M+N)-4}|\, r^{3(M+N)-4} \int_{\mathbb{R}^{3M}} \left[\frac{|\mathbb{S}^{3N-4}|}{|\mathbb{S}^{3(M+N)-4}|r^{3M}} \left(1-\frac{\mathbf{|t|}^2}{r^2}\right)_{+}^{\frac{3N-5}{2}}\Gamma^{-1/2}(\mathbf{t}) - \Gamma^{1/2}(\mathbf{t})\right]^2 d\mathbf{t} \, dr
\]
}

By the appropriate sequence of variable changes, the top expression can be shown to equal $\int_{\mathbb{R}^{3M}} \Gamma(\mathbf{v})(h(\mathbf{v})-1)^2d\mathbf{v} = \|h-1\|^2$. The bottom expression has been computed in \cite{bonetto}, except with $M$ and $N$ substituted where we have $3M$ and $3N-3$. The value computed was $M/(N-2)$, so in our case this is changed to $3M/(3N-5)$. Therefore:
\[
\|\mathbf{R}[h] - H\| \le \sqrt{\frac{3M}{3N-5}} \|h-1\|
\]

Now, we bound $\|H-1\|$. We rewrite $H$ as
\[
H(\vec{V}) = \int_{\mathbb{R}^{3}} \int_{\mathbb{R}^{3M-3}} h(\tilde{P}[\mathbf{u}, \vec{V}, \vec{Y}]) \, \Gamma(\mathbf{u}, \vec{Y}) \, d\mathbf{u} \, d\vec{Y},
\]
and now we perform another change of variables. If we define
\[
\vec{s} = \frac{\sqrt{M}\vec{V}+\sqrt{N}\vec{Y}}{\sqrt{M+N}},
\]
we can write $h(\tilde{P}[\mathbf{u}, \vec{V}, \vec{Y}]) = h(Q[\mathbf{u}, \vec{s}]) = h^*(\mathbf{u},\vec{s})$, defining $Q$ as:
\[
Q = [\mathbf{a_1},...,\mathbf{a_{3M-3}},\mathbf{e_1},\mathbf{e_2},\mathbf{e_3}],
\]
where we are now keeping all the vectors in $\mathbb{R}^{3M}$. Note that $Q$ is an orthogonal matrix.

We rewrite $H$, changing integration variables to use $\vec{s}$ instead of $\vec{Y}$. 
\[
H(\vec{V}) = \int_{\mathbb{R}^{3}} \int_{\mathbb{R}^{3M-3}} h^*(\mathbf{u}, \vec{s}) \, \Gamma(\mathbf{u}, \vec{Y}) \, d\mathbf{u} \, d\vec{Y} 
\]
\[
= \sqrt{\frac{M+N}{N}}^3\int_{\mathbb{R}^{3}} \int_{\mathbb{R}^{3M-3}} h^*(\mathbf{u}, \vec{s}) \, \Gamma\left(\mathbf{u}, \sqrt{\frac{M+N}{N}}\vec{s} - \sqrt{\frac{M}{N}}\vec{V}\right) \, d\mathbf{u} \, d\vec{s}
\]
\[
= \int_{\mathbb{R}^{3}} \int_{\mathbb{R}^{3M-3}} h^*(\mathbf{u},\vec{s})\, \Gamma(\mathbf{u},\vec{s}) \, n(\vec{s},\vec{V}) \, d\mathbf{u} \, d\vec{s},
\]
where
\[
n(\vec{s},\vec{V}) = \sqrt{\frac{M+N}{N}}^3\exp{\left[-\pi\left( \frac{M}{N}|\vec{s}|^2 + \frac{M}{N}|\vec{V}|^2 - 2\frac{\sqrt{M(M+N)}}{N}\vec{s} \cdot \vec{V} \right)\right]}
\]

Notice that $H(\vec{V})$ must be 1 when $h = 1$, which means
\[
\int_{\mathbb{R}^{3}} \int_{\mathbb{R}^{3M-3}} \Gamma(\mathbf{u},\vec{s}) \, n(\vec{s},\vec{V}) \, d\mathbf{u} \, d\vec{s} = 1
\]
We can therefore use the trick from the analysis of $\|\mathbf{R}[h]-H\|$, writing:
\[
H(\vec{V})-1 = \int_{\mathbb{R}^{3}} \int_{\mathbb{R}^{3M-3}} (h^*(\mathbf{u},\vec{s})-1)\, (n(\vec{s},\vec{V})-1) \, \Gamma(\mathbf{u},\vec{s})  \, d\mathbf{u} \, d\vec{s},
\]

\[
(H(\vec{V})-1)^2 \le \int_{\mathbb{R}^{3}} \int_{\mathbb{R}^{3M-3}} (h^*(\mathbf{u},\vec{s})-1)^2 \, \Gamma(\mathbf{u},\vec{s})  \, d\mathbf{u} \, d\vec{s}
\]
\[
\times \int_{\mathbb{R}^{3}} \int_{\mathbb{R}^{3M-3}} (n(\vec{s},\vec{V})-1)^2 \, \Gamma(\mathbf{u},\vec{s})  \, d\mathbf{u} \, d\vec{s}
\]
Remembering that $h^*(\mathbf{u}, \vec{s}) = h(Q[\mathbf{u},\vec{s}])$ and that $Q$ is orthogonal, we can perform the change of variables to show the top integral equals $\|h-1\|^2$. 

Now we write $\|H(\vec{V})-1\|^2$ and substitute:
\[
\|H(\vec{V})-1\|^2 = \int_{\mathbb{R}^3} (H(\vec{V})-1)^2 \, \Gamma(\vec{V}) \, d\vec{V}
\]
\[
\le \|h-1\|^2 \, \int_{\mathbb{R}^{3}} \int_{\mathbb{R}^{3}} \int_{\mathbb{R}^{3M-3}} (n(\vec{s},\vec{V})-1)^2 \, \Gamma(\mathbf{u},\vec{s}, \vec{V})  \, d\mathbf{u} \, d\vec{s} \, d\vec{V}
\]
\[
= \|h-1\|^2 \left( \int_{\mathbb{R}^{3}} \int_{\mathbb{R}^{3}} \int_{\mathbb{R}^{3M-3}} n(\vec{s},\vec{V})^2 \, \Gamma(\mathbf{u},\vec{s}, \vec{V})  \, d\mathbf{u} \, d\vec{s} \, d\vec{V} - 1 \right)
\]
where we invoke the normalization of $n$ for the last equality.

\[
\int_{\mathbb{R}^{3}} \int_{\mathbb{R}^{3}} \int_{\mathbb{R}^{3M-3}} n(\vec{s},\vec{V})^2 \, \Gamma(\mathbf{u},\vec{s}, \vec{V})  \, d\mathbf{u} \, d\vec{s} \, d\vec{V} 
\]
{\scriptsize
\[
= \left( \frac{M+N}{N} \right)^3 \int_{\mathbb{R}^{3}} \int_{\mathbb{R}^{3}}\exp{\left[-\pi\left( \frac{2M+N}{N}|\vec{s}|^2 + \frac{2M+N}{N}|\vec{V}|^2 - 4\frac{\sqrt{M(M+N)}}{N}\vec{s} \cdot \vec{V} \right)\right]} d\vec{s}d\vec{V}
\]
}

The integral in the above expression is the cube of a double-integrated Gaussian with a quadratic form in the exponent. Carrying out the computation gives the simple result
\[
\int_{\mathbb{R}^{3}} \int_{\mathbb{R}^{3}} \int_{\mathbb{R}^{3M-3}} n(\vec{s},\vec{V})^2 \, \Gamma(\mathbf{u},\vec{s}, \vec{V})  \, d\mathbf{u} \, d\vec{s} \, d\vec{V}  = \left( \frac{M+N}{N} \right)^3
\]

Therefore,
\[
\|H(\vec{V})-1\| \le \sqrt{\left( \frac{M+N}{N} \right)^3-1} \, \|h-1\|
\]

We now have bounds on both $\mathbf{R}[h]-H$ and $\|H-1\|$, and collecting terms proves the lemma. \end{proof}

\textbf{Lemma 2.}
\[
\left\| \frac{1}{N}\sum_{j=1}^N R_{ij}^Iu - T_iu \right\|^2 = \frac{1}{N}\big(\langle u, T_i u \rangle - \langle T_i u, T_i u \rangle \big)
\]
for any $u(\mathbf{v})$ in $L^2(\mathbb{R}^{3M},\Gamma)$.

\begin{proof}

\[
\left\| \frac{1}{N}\sum_{j=1}^N R_{ij}^Iu - T_iu \right\|^2 = \int_{\mathbb{R}^{3(M+N)}} 
\left( \frac{1}{N} \sum_{j=1}^N R_{ij}^I u - T_i u \right)^2 \, \Gamma(\mathbf{v}, \mathbf{w}) \, d\mathbf{v} \, d\mathbf{w}
\]

\[
= \frac{1}{N^2} \sum_{j,k=1}^N \int_{\mathbb{R}^{3(M+N)}} R_{ij}^I u \, R_{ik}^I u \, \Gamma(\mathbf{v}, \mathbf{w}) \, d\mathbf{v} \, d\mathbf{w}
\]

\[
- \frac{2}{N} \sum_{j=1}^N \int_{\mathbb{R}^{3(M+N)}} R_{ij}^I u \, T_i u \, \Gamma(\mathbf{v}, \mathbf{w}) \, d\mathbf{v} \, d\mathbf{w}
\]

\[
+ \int_{\mathbb{R}^{3(M+N)}} T_i u \, T_i u \, \Gamma(\mathbf{v}, \mathbf{w}) \, d\mathbf{v} \, d\mathbf{w}
\]

We analyze the integral in each of these terms, starting with the $\frac{1}{N^2}$ term. For \( j \ne k \): 
\[
\int_{\mathbb{R}^{3(M+N)}} R_{ij}^I u \, R_{ik}^I u \, \Gamma(\mathbf{v}, \mathbf{w}) \, d\mathbf{v} \, d\mathbf{w}
\]
\[
= \int_{\mathbb{R}^{3(M+N)}} \int_{\mathbb{S}^2} u\big(\mathbf{v}_{i}^*(\vec{\Omega_j})\big) \, d\sigma(\vec{\Omega_j}) \, \int_{\mathbb{S}^2} u\big(\mathbf{v}_{i}^*(\vec{\Omega_k})\big) \, d\sigma(\vec{\Omega_k}) \, \Gamma(\mathbf{v}, \mathbf{w}) \, d\mathbf{v} \, d\mathbf{w}
\]

\[
= \int_{\mathbb{R}^{3(M+N-2)}} \left( \int_{\mathbb{R}^3} \int_{\mathbb{S}^2} \Gamma(\vec{w}_j) \, u\big(\mathbf{v}_{i}^*(\vec{\Omega_j})\big) \, d\sigma(\vec{\Omega_j}) d\vec{w}_j \right)
\]
\[
\times \left( \int_{\mathbb{R}^3}\int_{\mathbb{S}^2} \Gamma(\vec{w}_k) \, u\big(\mathbf{v}_{i}^*(\vec{\Omega_k})\big) \, d\sigma(\vec{\Omega_k}) d\vec{w}_k \, \right) \Gamma(\mathbf{v}, \mathbf{w}^{jk}) \, d\mathbf{v} \, d\mathbf{w}^{jk}
\]

\[
= \int_{\mathbb{R}^{3(M+N-2)}} T_i u \, T_i u \, \Gamma(\mathbf{v}, \mathbf{w}^{jk}) \, d\mathbf{v} \, d\mathbf{w}^{jk}
\]

\[
= \int_{\mathbb{R}^{3(M+N)}} T_i u \, T_i u \, \Gamma(\mathbf{v}, \mathbf{w}) \, d\mathbf{v} \, d\mathbf{w} = \langle T_i u, T_i u \rangle
\]

For \( j = k\):
\[
\int_{\mathbb{R}^{3(M+N)}} R_{ij}^I u \, R_{ij}^I u \, \Gamma(\mathbf{v}, \mathbf{w}) \, d\mathbf{v} \, d\mathbf{w}
= \int_{\mathbb{R}^{3(M+N)}} u \, R_{ij}^I u \, \Gamma(\mathbf{v}, \mathbf{w}) \, d\mathbf{v} \, d\mathbf{w},
\]
and by a similar analysis as the \( j \ne k\) case,
\[
= \int_{\mathbb{R}^{3(M+N)}} u \, T_i u \, \Gamma(\mathbf{v}, \mathbf{w}) \, d\mathbf{v} \, d\mathbf{w} = \langle u, T_i u \rangle
\]
Finally,
\[
\int_{\mathbb{R}^{3(M+N)}} R_{ij}^I u \, T_i u \, \Gamma(\mathbf{v}, \mathbf{w}) \, d\mathbf{v} \, d\mathbf{w}
\]
\[
= \int_{\mathbb{R}^{3(M+N)}} T_i u \, T_i u \, \Gamma(\mathbf{v}, \mathbf{w}) \, d\mathbf{v} \, d\mathbf{w}
= \langle T_i u, T_i u \rangle
\]
Combining terms, we get:
\[
\left\| \frac{1}{N}\sum_{j=1}^N R_{ij}^Iu - T_iu \right\|^2 = \frac{1}{N^2}\big[(N^2-N)\langle T_i u, T_i u \rangle
+ N\langle u, T_i u \rangle\big]
\]

\[
- \frac{2}{N}N\langle T_i u, T_i u \rangle + \langle T_i u, T_i u \rangle
\]

\[
= \frac{1}{N}\big(\langle u, T_i u \rangle - \langle T_i u, T_i u \rangle \big)
\]

\end{proof}

\textbf{Lemma 3.}
For any function $u(\mathbf{v})$ in $L^2(\mathbb{R}^{3M},\Gamma)$ satisfying $\langle u, 1 \rangle = 0$,
\[
\langle u, T_i[u] \rangle \le \frac{2}{3}\langle u,u \rangle
\]
for some $i$ from 1 to $M$.

\begin{proof}

Because the set of polynomials is dense in $L^2(\mathbb{R}^{3M},\Gamma)$, it suffices to show $\langle p, T_i[p] \rangle \le \frac{2}{3}\langle p,p \rangle$ for any polynomial $p$ satisfying $\langle p, 1 \rangle = 0$. 

Moreover, $T_i$ acts as the identity on functions which do not depend on $\vec{v}_i$. Therefore, it will suffice to prove the statement for $i = 1$ and polynomials of only $\vec{v}_1$. For simplicity, we will write $T_1 = T$ and $\vec{v}_1 = \vec{v} = (v_1, v_2, v_3)$. 

Let $H_{m}(x)$ be the Hermite polynomial in $x$ of degree $m$. We define $H_{\vec{m}}(\vec{x}) = H_{m_1}(x_1)H_{m_2}(x_2)H_{m_3}(x_3)$ to be the Hermite polynomial in $\vec{x}$ of total degree $|\vec{m}| = m_1 + m_2 + m_3$. 

Because $T$ is self-adjoint, we write:
\[
\langle T[H_{\vec{\alpha}}(\vec{v})], H_{\vec{n}}(\vec{v})\rangle = \langle H_{\vec{\alpha}}(\vec{v}), T[H_{\vec{n}}(\vec{v})]\rangle
\]
It is straightforward to see that $T$ preserves total degree of polynomials. As a result, if $|\vec{\alpha}| < |\vec{n}|$, the left inner product will be 0 by orthogonality of the Hermite polynomials. For the same reason, if $|\vec{\alpha}| > |\vec{n}|$, the right inner product will be 0. Therefore, defining the space $V_m = \text{span}\{ H_{\vec{m}} : |\vec{m}| = m\}$, we can see that $T$ maps $V_m$ to $V_m$.

An arbitrary polynomial $p(\vec{v})$ in $V_m$ can be written as:
\[
p(\vec{v}) = \sum_{|\vec{m}| \le m}a_{\vec{m}}v^{m_1}_1v^{m_2}_2v^{m_3}_3
\]

We define the operator $J$ such that:
\[
J[p](\vec{v}) = \sum_{|\vec{m}| = m}a_{\vec{m}}v^{m_1}_1v^{m_2}_2v^{m_3}_3
\]
$J$ is a map from $V_m$ to the space of homogeneous polynomials of total degree $|\vec{m}|$, which we call $W_m$. $J$ is linear and one-to-one, and we can thus define the operator $\tilde{T} = JTJ^{-1}$, which maps $W_m$ to $W_m$. 

A polynomial $q(\vec{v})$ in $W_m$ can be written as:
\[
q(\vec{v}) = \sum_{|\vec{m}| \le m}a_{\vec{m}}v^{m_1}_1v^{m_2}_2v^{m_3}_3 = \prod^{m}_{k=1}(\vec{c}_k, \vec{v})
\]
Written in the latter form, we can identify $q$ in $W_m$ with $S$ in $(\mathbb{R}^3)^{\otimes m}_{\text{sym}}$, where we define
\[
S = \bigotimes_{k=1}^m \vec{c}_k
\]

Using the definitions of $T$ and $\tilde{T}$, we can describe the action of $\tilde{T}$ on $S$ as:
\[
\tilde{T}[S] = \int_{\mathbb{R}^3} \bigotimes_{k=1}^m (I-\Omega \otimes \Omega)\vec{c}_k \, d\vec{\Omega} = \int_{\mathbb{R}^3} A^{\otimes m}_{\vec{\Omega}} S \, d{\vec{\Omega}},
\]
where $A_{\vec{\Omega}} = I - \Omega \otimes \Omega$.

Now, we use this setup to write:
\[
\sup_{p \in V_m, \langle p,1\rangle=0}\frac{\langle p,Tp\rangle}{\langle p,p\rangle}
= \sup_{q \in W_m, \langle q,1\rangle=0}\frac{\langle q,\tilde{T}q\rangle}{\langle q,q\rangle}
\le \sup_{S \in (\mathbb{R}^3)^{\otimes m}, (S,1)=0}\frac{(S,\tilde{T}S)}{(S,S)},
\]
where we exploit the fact that an operator's spectral gap does not depend on our choice of inner product. The inner product denoted above as $(f,g)$ is the standard inner product defined on a tensor product space of vectors in $\mathbb{R}^3$. The inequality in the last step comes from the fact that $(\mathbb{R}^3)^{\otimes m}_{\text{sym}}$ is a subset of $(\mathbb{R}^3)^{\otimes m}$. 

Finally, we write out:
\[
\sup_{S \in (\mathbb{R}^3)^{\otimes m}, (S,1)=0}\frac{(S,\tilde{T}S)}{(S,S)} = \sup_{S \in (\mathbb{R}^3)^{\otimes m}, (S,1)=0}\frac{\int_{\mathbb{R}^3} (S, A^{\otimes m}_{\vec{\Omega}}S) \, d{\vec{\Omega}}}{(S,S)},
\]
and a direct computation shows the quantity on the right $\le 2/3$, thus completing the proof. \end{proof}

\section{Main Result}

\textbf{Theorem.}
Let the initial state of the system be the $L^2(\mathbb{R}^{3M},\Gamma)$ function $h_0(\mathbf{v})$, satisfying $\langle h, 1 \rangle = 1$,
\[
\| h_t - \tilde{h}_t \| \le \left[C(\frac{M}{N})\big(1 - e^{-\lambda Mt}\big)+ b\frac{M}{\sqrt{N}}\big(e^{-\mu t/3} - e^{-kt} \big)  \right] \|h_0-1\|,
\]
where:
\begin{itemize}
    \item $C(\frac{M}{N}) = \sqrt{\frac{3M}{3N-5}}+\sqrt{(\frac{M+N}{N})^3-1}$. The $\frac{M}{N}$ "argument" of $C$ indicates that the scaling of $C$ is of order $M/N$.
    \item $\lambda = \lambda_S/2+\mu$.
    \item $k$ and $b$ are constants independent of $M$ and $N$; $k$ is positive and $b$ has the same sign as $k - \mu/3$.
\end{itemize}

Thus, the bound initially increases rapidly from 0 and has a "bump" of order $M/\sqrt{N}$, but the long-term behavior is of order $M/N$.

\begin{proof}

We start by writing $h_0 = u_0 + 1$, where $\langle u_0, 1 \rangle = 0$. Since 1 is a steady state for both evolutions, we have:
\[
h_t - \tilde{h}_t = \big( e^{\mathcal{L}t} - e^{\bar{\mathcal{L}}t} \big) u_0
\]

\[
= \int_0^t e^{\mathcal{L}(t-s)} (\mathcal{L} - \bar{\mathcal{L}}) e^{\bar{\mathcal{L}}s} u_0 \, ds,
\]
invoking the Duhamel expansion. Applying the $L^2$ norm, we get:
\[
\| h_t - \tilde{h}_t \| 
= \left\| \int_0^t e^{\mathcal{L}(t-s)} (\mathcal{L} - \bar{\mathcal{L}}) e^{\bar{\mathcal{L}}s} u_0 \, ds \right\|.
\]

Our strategy will be to use the operator $\mathbf{R}$ to split the above expression into two parts. For the first part, we will exploit rotational invariance to produce the permanent bound of order $M/N$. Then we will follow a very similar argument to that of \cite{bonetto} to show the second part has a temporary bound of order $M/\sqrt{N}$. 

We split the expression by writing
{\small
\[
\int_0^t e^{\mathcal{L}(t-s)} (\mathcal{L} - \bar{\mathcal{L}}) e^{\bar{\mathcal{L}}s} u_0 \, ds = \int_0^t e^{\mathcal{L}(t-s)} \mathbf{R}(\mathcal{L} - \bar{\mathcal{L}}) e^{\bar{\mathcal{L}}s} u_0 \, ds + \int_0^t e^{\mathcal{L}(t-s)} (I-\mathbf{R})(\mathcal{L} - \bar{\mathcal{L}}) e^{\bar{\mathcal{L}}s} u_0 \, ds,
\]
}
and, invoking the triangle inequality,
{\small
\[
\| h_t - \tilde{h}_t \| \le \left\| \int_0^t e^{\mathcal{L}(t-s)} \mathbf{R}(\mathcal{L} - \bar{\mathcal{L}}) e^{\bar{\mathcal{L}}s} u_0 \, ds \right\| + \left\| \int_0^t e^{\mathcal{L}(t-s)} (I-\mathbf{R})(\mathcal{L} - \bar{\mathcal{L}}) e^{\bar{\mathcal{L}}s} u_0 \, ds \right\|.
\]
}

Considering the first expression, we remember $\mathbf{R}$ outputs functions which are invariant under all momentum-preserving rotations, meaning all $R_{ij}^x$, for $x=S,R,I$. Therefore, $\mathcal{L}$ acts as the identity on outputs of $\mathbf{R}$, so we write:
\[
\left\| \int_0^t e^{\mathcal{L}(t-s)} \mathbf{R}(\mathcal{L} - \bar{\mathcal{L}}) e^{\bar{\mathcal{L}}s} u_0 \, ds \right\| = \left\| \int_0^t  \mathbf{R}(\mathcal{L} - \bar{\mathcal{L}}) e^{\bar{\mathcal{L}}s} u_0 \, ds \right\|
\]

Next, we note that $\mathbf{R}$ averages over all the states that each $R_{ij}^x$ averages over, meaning the two-particle averages can be absorbed into the general average: $\mathbf{R}R_{ij}^x = \mathbf{R}$. Remembering the form of $\mathcal{L}$, we can see from this that $\mathbf{R}\mathcal{L} = 0$. Thus:
\[
\left\| \int_0^t  \mathbf{R}(\mathcal{L} - \bar{\mathcal{L}}) e^{\bar{\mathcal{L}}s} u_0 \, ds \right\| = 
\left\| \int_0^t  -\mathbf{R} \bar{\mathcal{L}} e^{\bar{\mathcal{L}}s}u_0  \, ds \right\|
\]
\[
\left\| -\mathbf{R} \left( \int_0^t  \bar{\mathcal{L}} e^{\bar{\mathcal{L}}s}  \, ds \right) u_0\right\| = \left\| \mathbf{R}(I-e^{\bar{\mathcal{L}}t})u_0 \right\| = \left\| \mathbf{R}[g] - 1 \right\|,
\]
where $g = (I-e^{\bar{\mathcal{L}}t})u_0 + 1$. Since $\bar{\mathcal{L}}$ does not involve the reservoir particles when acting on a function of $\mathbf{v}$ only, $g$ also depends only on $\mathbf{v}$. Furthermore, by the self-adjointness of $\bar{\mathcal{L}}$, $\langle g,1 \rangle = \langle u_0+1, 1 \rangle = \langle h_0, 1 \rangle=1$. Thus, we can apply Lemma 1 to $\left\| \mathbf{R}[g] - 1 \right\|$ to get:
\[
\left\| \mathbf{R}[g] - 1 \right\| \le C(\frac{M}{N}) \left\| g-1 \right\| = C(\frac{M}{N})\left\| (I-e^{\bar{\mathcal{L}}t})u_0 \right\|.
\]

Now we need to maximize $\left\| (I-e^{\bar{\mathcal{L}}t})u_0 \right\|$; this is equivalent to minimizing $\langle u_0, \bar{\mathcal{L}}u_0 \rangle$. When considering the action of $\bar{\mathcal{L}}$ on $u_0$, the action of $\mathcal{L}_R$ is trivial, again because $u_0$ depends only on $\mathbf{v}$. 

Looking at $\mathcal{L}_S$ and $\mathcal{L}_T$, each term has an $(R^S_{ij} - I)$ or a $(T_i - I)$; the infimum of the eigenvalues of each of these is -1. $\mathcal{L}_S$ is a sum of $M(M-1)/2$ terms divided by $M-1$, and $\mathcal{L}_T$ is a sum of $M$ terms, so the infimum of the eigenvalues of $\mathcal{L}_S + \mathcal{L}_T$ is the negative constant $-\lambda M$, where $\lambda = \lambda_S/2+\mu$. Thus, $\langle u_0, \bar{\mathcal{L}}u_0 \rangle \le -\lambda M \langle u_0, u_0 \rangle$, and we write:
\[
\left\| (I-e^{\bar{\mathcal{L}}t})u_0 \right\| \le (1-e^{\lambda Mt})\|u_0\| = (1-e^{\lambda Mt})\|h_0-1\|
\]
Therefore:
\[
\left\| \int_0^t e^{\mathcal{L}(t-s)} \mathbf{R}(\mathcal{L} - \bar{\mathcal{L}}) e^{\bar{\mathcal{L}}s} u_0 \, ds \right\| \le C(\frac{M}{N})(1-e^{\lambda Mt})\|h_0-1\|
\]

This takes care of the first expression. Moving on to the second expression, we write:
\[
\left\| \int_0^t e^{\mathcal{L}(t-s)} (I-\mathbf{R})(\mathcal{L} - \bar{\mathcal{L}}) e^{\bar{\mathcal{L}}s} u_0 \, ds \right\|
\le \int_0^t \left\| e^{\mathcal{L}(t-s)} (I-\mathbf{R})(\mathcal{L} - \bar{\mathcal{L}}) e^{\bar{\mathcal{L}}s} u_0 \right\|ds
\]
It has been proven in \cite{loss} that the 3D Kac evolution operator, in our case $\mathcal{L}$ for the system and reservoir, has a spectral gap. The largest eigenvalue is 0 for functions invariant under momentum-preserving rotations, and the second-largest is a constant $-k$ where $k > 0$. $I-\mathbf{R}$ is a projector onto the orthogonal space to functions invariant under momentum-preserving rotations, so any function output by $I-\mathbf{R}$ will have an eigenvalue less than $-k$:
\[
\int_0^t \left\| e^{\mathcal{L}(t-s)} (I-\mathbf{R})(\mathcal{L} - \bar{\mathcal{L}}) e^{\bar{\mathcal{L}}s} u_0 \right\|ds \le 
\int_0^t  e^{-k(t-s)} \left\|(I-\mathbf{R})(\mathcal{L} - \bar{\mathcal{L}}) e^{\bar{\mathcal{L}}s} u_0 \right\|ds
\]
\[
\le \int_0^t  e^{-k(t-s)} \left\|(\mathcal{L} - \bar{\mathcal{L}}) e^{\bar{\mathcal{L}}s} u_0 \right\|ds,
\]
where for the last line we again take advantage of the fact that $I-\mathbf{R}$ is an orthogonal projector.

Moving on, we note that $u_s(\mathbf{v}) = e^{\bar{\mathcal{L}}s} [u_0](\mathbf{v})$ depends only on $\mathbf{v}$. Expanding out $\mathcal{L} - \bar{\mathcal{L}}$ and invoking the Cauchy-Schwarz inequality, we have:
\[
\left\| (\mathcal{L} - \bar{\mathcal{L}} )u_s \right\|^2
= \left\| \mu \sum_{i=1}^M \big(\frac{1}{N}\sum_{j=1}^N R_{ij}^Iu_s - T_iu_s\big) \right\|^2
\le \mu^2 M \sum_{i=1}^M \left\| \frac{1}{N}\sum_{j=1}^N R_{ij}^Iu_s - T_iu_s \right\|^2
\]
Now we can use Lemma 2 to write:
\[
\mu^2 M \sum_{i=1}^M \left\| \frac{1}{N}\sum_{j=1}^N R_{ij}^Iu_s - T_iu_s \right\|^2 \le
\mu^2 \frac{M}{N} \sum_{i=1}^M \big(\langle u_s, T_i u_s \rangle - \langle T_i u_s, T_i u_s \rangle \big)
\le \mu^2 \frac{M^2l^2}{N}\| u_s \|^2
\]
where we define $l^2 = \sup_{\|u\| = 1}\big(\langle u, T_i u \rangle - \langle T_i u, T_i u \rangle \big)$. So now we have:
\[
\left\| \int_0^t e^{\mathcal{L}(t-s)} (I-\mathbf{R})(\mathcal{L} - \bar{\mathcal{L}}) e^{\bar{\mathcal{L}}s} u_0 \, ds \right\|
\le \int_0^t  e^{-k(t-s)} \mu l \frac{M}{\sqrt{N}} \left\| e^{\bar{\mathcal{L}}s} u_0 \right\|ds
\]

Finally, we need to consider $\left\| e^{\bar{\mathcal{L}}s} u_0 \right\|$. We have:
\[
\langle u, \bar{\mathcal{L}} u \rangle = \langle u, \mathcal{L}_S u \rangle + \langle u, \mathcal{L}_R u \rangle + \langle u, \mathcal{L}_T u \rangle \le \langle u, \mathcal{L}_T u \rangle
\]

\[
= \langle u,\mu \sum_{i=1}^M(T_iu-u) \rangle = \mu \big(\sum_{i=1}^M\langle u, T_iu \rangle - M \langle u, u \rangle \big)
\]

Our goal is to prove:
\[
\left\| e^{\bar{\mathcal{L}}s} u_0 \right\| \le e^{-\mu\kappa s} \|u_0\|, \quad \text{where } \kappa > 0
\]

To do this, it suffices to show
\[
\mu \big(\sum_{i=1}^M\langle u_0, T_iu_0 \rangle - M \langle u_0, u_0 \rangle \big) \le -\mu\kappa \langle u_0,u_0 \rangle
\]

By Lemma 3, because $u_0$ is a function in $L^2(\mathbb{R}^{3M},\Gamma)$ satisfying $\langle u_0, 1 \rangle = 0$, we know $\langle u_0, T_i[u_0] \rangle \le \frac{2}{3}\langle u_0,u_0 \rangle$ for some $i$ from 1 to $M$. We also know $\langle u_0, T_i[u_0] \rangle \le \frac{2}{3}\langle u_0,u_0 \rangle$ for any function $u_0$ in $L^2(\mathbb{R}^{3M},\Gamma)$ (the computation can be carried out straightforwardly using the Cauchy-Schwarz inequality). Therefore:
\[
\mu \big(\sum_{i=1}^M\langle u_0, T_iu_0 \rangle - M \langle u_0, u_0 \rangle \big) \le -\mu/3 \langle u_0,u_0 \rangle,
\]
and so,
\[
\left\| e^{\bar{\mathcal{L}}s} u_0 \right\| \le e^{-\mu s/3} \|u_0\| = e^{-\mu s/3} \|h_0 - 1\|
\]

Putting it all together for the second expression gives:
\[
\left\| \int_0^t e^{\mathcal{L}(t-s)} (I-\mathbf{R})(\mathcal{L} - \bar{\mathcal{L}}) e^{\bar{\mathcal{L}}s} u_0 \, ds \right\|
\le \int_0^t  e^{-k(t-s)} \mu l \frac{M}{\sqrt{N}} e^{-\mu s/3} \|h_0 - 1\|ds
\]
\[
= b\frac{M}{\sqrt{N}}\big(e^{-\mu t/3} - e^{-kt} \big)\|h_0 - 1\|,
\]
where $b = \frac{l}{k-\mu/3}$.

Adding the first and second expressions gives our final result:
\[
\boxed{
\| h_t - \tilde{h}_t \| \le \left[C(\frac{M}{N})\big(1 - e^{-\lambda Mt}\big)+ b\frac{M}{\sqrt{N}}\big(e^{-\mu t/3} - e^{-kt} \big)  \right] \|h_0-1\|
}
\]
\end{proof}

\bibliographystyle{IEEEtran}
\bibliography{references}

\end{document}